\documentstyle[aps,preprint,epsf]{revtex} \bibstyle{unsrt}
\newcommand {\ket}[1]{|\,{#1}\,\rangle}
\newcommand {\bra}[1]{\langle\,{#1}\,|}

\begin{document}
\draft
\title{Coherently Controlled Nanoscale Molecular Deposition}

\author{Bijoy K. Dey, Moshe Shapiro$^*$ and Paul Brumer} \address{Chemical Physics Theory Group, Department  of Chemistry, University of Toronto, Toronto, Ontario M5S 3H6, Canada} 

\date{\today} \maketitle 

\begin{abstract}

Quantum interference effects are shown to provide a means of controlling and enhancing the focusing of a collimated neutral molecular beam onto a surface. The nature of the aperiodic pattern formed can be altered by varying laser field characteristics and the system geometry.
\end{abstract}

\vspace{1.5in}
\noindent
$^*$Permanent Address: Chemical Physics Department, The Weizmann Institute of Science, Rehovot, Israel

\vspace{0.2in}

\noindent
Light induced forces have been used to deflect, slow, accelerate, cool, and confine \cite{mechanical} neutral atoms. Similarly, and of particular interest here, atoms have been focused and deposited on surfaces on the nanometer scale\cite{focus,focus2}. In these cases, preliminary laser cooling followed by passage through an optical standing wave resulted in the formation of a periodic submicron atomic patterns on a surface. There are far fewer results for molecules, the most noteworthy being experiments\cite{corkum} and theory\cite{CMmol} on focusing of molecules using intense laser fields.  In this letter we show that coherent control techniques\cite{cc} can be used to enhance and control the deposition of  molecules on a surface, in {\em aperiodic}, nanometric scale patterns. The essence of the technique lies in the pre-preparation of an initial controlled superposition of molecular eigenstates, followed by passage through an optical standing wave composed of two fields of related frequency. By varying the characteristics of the prepared superposition, or the characteristics of the optical standing wave, one can vary the induced dipole-electric field interaction, and hence alter the deposited pattern. The result demonstrates the utility of coherent control to manipulate the translational motion of molecules. Below we describe the general theory and provide computational results for the computationally convenient\cite{mccullough} molecule, N$_2$.

\begin{figure}[!h]
\epsfxsize=6.0in
\hspace*{-2.0cm}\epsffile{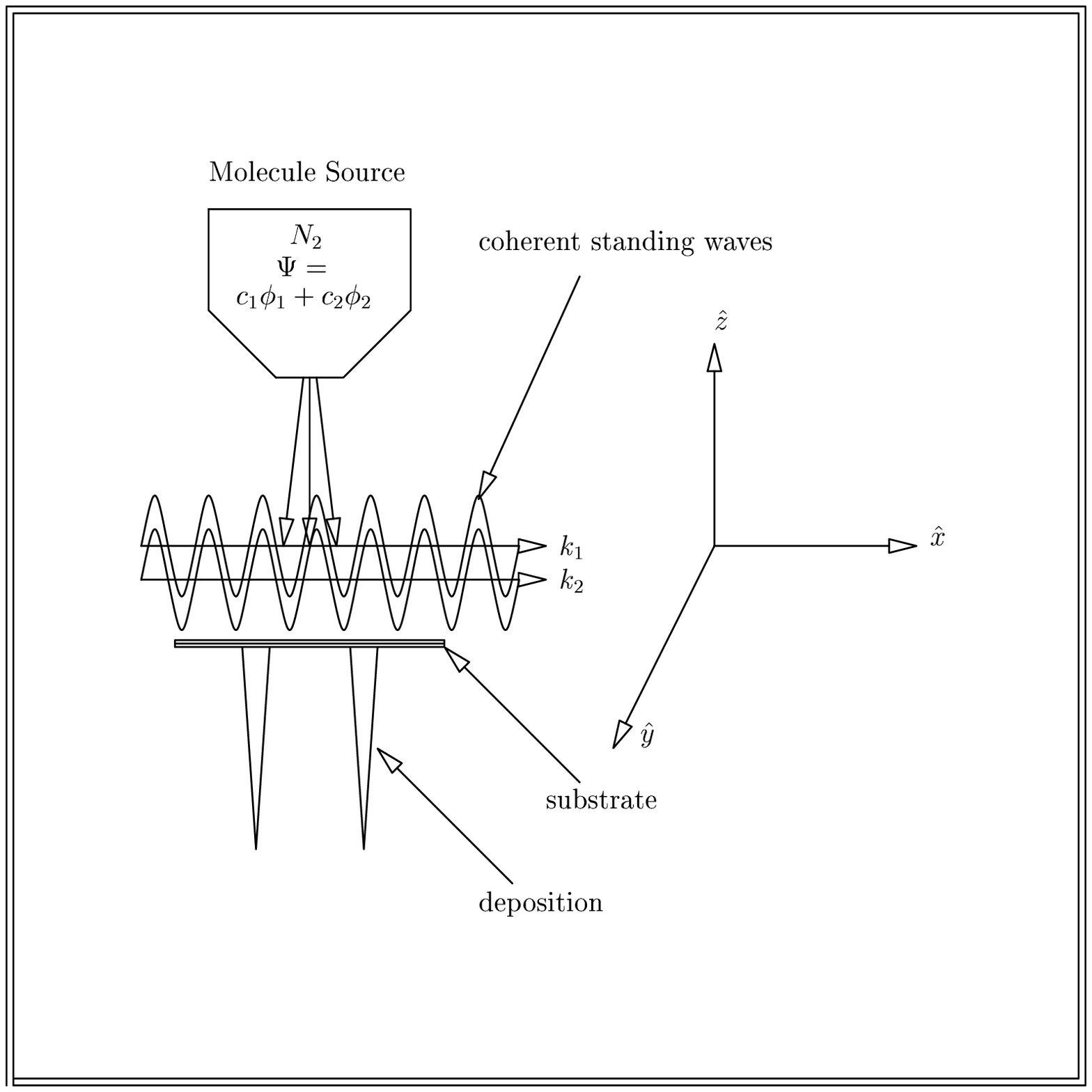}
\vspace*{-4.0cm}\caption{Schematic of proposed control scenario.}
\label{fig1}
\end{figure}

The general configuration of the proposed control scenario is illustrated in  Fig. \ref{fig1}. A beam of neutral molecules propagating along the $z$ direction, is prepared in a superposition of vib-rotational states with a highly cooled transverse velocity distribution. Preparation is achieved either by passing a pre-cooled beam through a preparatory electric field to create a superposition of vib-rotational states in the ground electronic state, or by simultaneously preparing the superposition and cooling the transverse velocity by an extension of a recently proposed radiative association approach\cite{vardi}. For simplicity we focus on a two-level superposition, i.e. 

\begin{eqnarray} \label{superposition}
\ket{\Psi (t)}=c_1\ket{\phi _1}e ^{-iE_1t/\hbar}+c_2\ket{\phi _2}e^{-iE_2t/\hbar} 
\end{eqnarray}
where  $\ket{\phi _i}$ are the eigenstates of the molecular Hamiltonian, of energy $E_i$. 

The molecules then pass through a standing wave ${\bf E}(x,t)$ composed of two electromagnetic fields which lie parallel to the surface and are polarized in the $z$ direction. That is, 
\begin{eqnarray} \label{Efield}
{\bf E}(x,t)
&=& [2 E_1^{(0)}\cos(k_1x)e^{i\omega_1 t} + cc~~]~ {\hat{\bf k}}+
[ 2 E_2^{(0)}\cos(k_2 x + \theta_F)e^{i\omega_2 t} +cc~~ ]~ {\hat{\bf k}}+
\\\nonumber
&\equiv& [E(\omega_1) + cc~~]~ {\hat{\bf k}} + [E(\omega_2) + cc~~]~ {\hat{\bf k}}
\end{eqnarray}
Here  ${\hat{\bf k}}$ denotes a unit vector in the $z$ direction, $cc$ denotes the complex conjugate of the terms preceding it, $\theta _F$ is the relative  phase of the two standing waves (SW), $E_j^{(0)}$, $k_j$,  and $\omega   _j$  are  the maximum amplitude, wave vector, and frequency of the jth standing wave, of wavelength $\lambda_j$.

The potential energy of interaction $V(x)$ of the molecule with the field ${\bf E}(x,t)$ is $V(x) = -\mu^{ind} \cdot {\bf E}(x,t)$, where $\mu^{ind}$ is the induced dipole moment. Within first order perturbation theory, the induced dipole of the superposition state in the presence of the two fields, chosen so that $E_1 + \hbar \omega_1 = E_2 + \hbar \omega_2$  is given by\cite{mccullough}:

\begin{eqnarray} \label{dipole}
{\bf  \mu}  ^{ind}&=&\chi  ^{in}(\omega   _1)E(\omega     _1)+\chi
^{ni}(\omega   _1)E(\omega   _1)+\chi   ^{in}(\omega
_2)E(\omega _2)+\chi  ^{ni}(\omega  _2)E(\omega _2)\\\nonumber
&&+\chi ^{in}(\omega   _{21}+\omega      _1)E(\omega    _{21}+\omega
_1)+\chi  ^{in}(\omega    _{21}-\omega   _2)E(\omega_{21}-\omega _2) + cc 
\end{eqnarray}
where $E(\omega_{21}+\omega_1)=2E_1^{(0)}\cos(k_1x)e^{i(\omega_{21}+\omega_1)t}$ and $E(\omega_{21}-\omega_2)=2E_2^{(0)}\cos(k_2x+\theta_F)e^{i(\omega_{21}-\omega_2)t}$. The $\chi$ are  the  following contributions to the $zz$ component of the polarizability:
\begin{eqnarray*}
\chi   ^{in}(\omega   _1)=\frac{1}{\hbar}\sum_j   c_1c_2^*[\frac{\mu
_{j1}^z\mu    _{2j}^z}{\omega_{j1}+\omega     _2}+\frac{\mu
_{j2}^z\mu      _{1j}^z}{\omega  _{j2}-\omega_2}]\frac{E_2^{(0)}}{E_1^{(0)}} 
\end{eqnarray*}
\begin{eqnarray*}
\chi   ^{in}(\omega    _2)&=&\frac{1}{\hbar}\sum_j  c_2c_1^*[\frac{\mu
_{j2}^{z}\mu     _{1j}^z}{\omega_{j2}+\omega    _1}+\frac{\mu
_{j1}^{z}\mu              _{2j}^z}{\omega         _{j1}-\omega
_1}]\frac{E_1^{(0)}}{E_2^{(0)}} 
\end{eqnarray*}
\begin{eqnarray*}
\chi ^{ni}(\omega _1)&=&\frac{1}{\hbar}\sum_j\sum_{i=1,2} |c_i|^2\mu _{ji}^{z
}\mu      _{ij}^z[\frac{1}{\omega _{ji}+\omega   _1}+\frac{1}{\omega
_{ji}-\omega  _1}]
\end{eqnarray*}
\begin{eqnarray*}
\chi ^{ni}(\omega _2)&=&\frac{1}{\hbar}\sum_j \sum_{i=1,2}|c_i|^2\mu _{ji}^{
z}\mu   _{ij}^z[\frac{1}{\omega     _{ji}+\omega  _2}+\frac{1}{\omega
_{ji}-\omega  _2}]
\end{eqnarray*}
\begin{eqnarray*}
\chi       ^{in}(\omega     _{21}+\omega   _1)=\frac{1}{\hbar}\sum_j [
c_1c_2^*\frac{\mu      _{j1}^{z}\mu _{2j}^z}{\omega _{j1}+\omega
_1}+c_1^*c_2\frac{\mu _{j2}^{z}\mu  _{1j}^z}{\omega _{j2}-\omega_1} ]
\end{eqnarray*}
\begin{eqnarray*}
\chi   ^{in}(\omega    _{21}-\omega        _2)=\frac{1}{\hbar}\sum_j [
c_1c_2^*\frac{\mu   _{j1}^{z}\mu  _{2j}^z}{\omega   _{j1}-\omega
_2}+c_1^*c_2\frac{\mu  _{j2}^{z}\mu _{1j}^z}{\omega _{j2}+\omega_2}] 
\end{eqnarray*}
where $\omega_{ij} = (E_j-E_i)/\hbar$ and $\mu_{ij}^z = \bra{\phi_i}\mu\cdot {\hat{\bf k}}\ket{\phi_j}$. Here the superscripts $``in"$ and $``ni"$  refer to the interference and non-interference contributions to  $\chi$, the interference terms being the direct consequence of the established coherence between $\ket{\phi_1}$ and $\ket{\phi _2}$ [Eq. (\ref{superposition})]. The summation in the above equations runs over  all  the vibrational    and  rotational states. For example, in the particular case of $N_2$, examined below,  vibrotational states of six excited electronic states ($b^    \prime$$^1$$\sum    _u^+$, $c^\prime$$^1$$\sum _u^+$,  $e^\prime$$^1$$\sum _u^+$,  $b^1\Pi   _u$, $c^1\Pi _u$ and $o^1\Pi _u$) are included. 

The final expression for the  potential within the rotating wave approximation
is then given by 
\begin{eqnarray} \label{pottot}
V(x)=-\mu^{ind}\cdot{\bf E}(x,t)= V^{ni}(x)+V^{in}(x) 
\end{eqnarray}
where 
\begin{eqnarray} \label{potni}
-V^{ni}(x)&=&2[4E_1^{(0){^2}}\cos      ^2(k_1x)\chi      ^{ni}(\omega
_1)+4E_2^{(0){^2}}\cos^2(k_2x+\theta _F)\chi^{ni}(\omega_2)\\\nonumber
&&+4E_1^{(0)}E_2^{(0)}\cos(k_1x)\cos(k_2x+\theta   _F)
[\chi^{ni}(\omega _1)+\chi^{ni}(\omega _2)]\cos(\omega_1-\omega_2)t
\end{eqnarray}
and 
\begin{eqnarray} \label{potin}
-V^{in}(x)&=&2[4E_1^{(0){^2}}\cos(k_1x)\cos(k_2x+\theta        _F)\chi
^{in}_r(\omega                                        _1)
+4E_2^{(0){^2}}\cos(k_1x)\cos(k_2x+\theta      _F)\chi  ^{in}_r(\omega
_2)\\\nonumber &&+4E_1^{(0)}E_2^{(0)}\cos^2(k_2x +   \theta  _F)[\chi
^{in}_r(\omega _1)\cos(\omega  _1-\omega  _2)t-\chi
^{in}_I(\omega     _1)\sin(\omega    _1-\omega  _2)t]\\\nonumber &&+
4E_1^{(0)}E_2^{(0)}\cos^2(k_1x)[\chi  ^{in}_r(\omega  _2)\cos(\omega
_2-\omega _1)t-\chi _I^{in}(\omega _2)  \sin(\omega_2-\omega _1)t]] 
\end{eqnarray}

Here  $\chi^{in}_r$ and $\chi^{in}_I$ denote the  real and imaginary part of the $zz$ component of  $\chi^{in}$. Computations show that the time dependent contributions to Eqs. (\ref{potni}) and  (\ref{potin}) average out and may be neglected. The resultant time independent ``optical potential" displays a series of maxima and minima along $x$, with each minima serving to focus the molecules, and each maxima serving to defocus them. The structure of $V(x)$ and hence its effect on the molecule's dynamics depends upon the {\em control parameters} $E_1^{(0)}$,  $E_2^{(0)}$, $c_1$,  $c_2$, $\theta _F$   and the quantum numbers $\nu_i, J,M$.

The extent to which control is possible is evident from the computational results shown below on $N_2$ (a molecule chosen solely for computational convenience). Here $\ket{\phi_i} \equiv \ket{\nu_i,J_i,M_i}$, where $\nu_i$ and $J_i$    are  the vibrational    and  rotational  quantum  numbers respectively and $M_i$ is the projection of  $J_i$ along the $z$ direction. Selection rules imply \cite{mccullough} that $\chi^{in}$ is zero unless $\ket{\phi_1}$ and $\ket{\phi_2}$ are of the same parity. To this end we employ a two photon preparatory step so that $J_2=J_1+2$, $M_1=M_2$.

As an example, we compute classical trajectories for the deposition of $N_2$ on a surface, reported as the number of trajectories $N(x)$ incident on the surface in a $\Delta x$ interval of 1.403 nm. Our initial studies examined deposition using a nozzle width of 20 $\mu m$ and a similar sample size. Results for the chosen parameters ($\lambda_1=  0.628 \mu m, \lambda_2 = 0.736 \mu m$) showed an almost periodic repeating  patterns  of  3-4  $\mu  m$ width. Hence we here focus down to this subregion, with computations simplified by reducing the nozzle diameter and sample size to $4 \lambda_2= 3 \mu m$. The initial velocity along $z$ is taken as 600 m/sec and the transverse velocity is assumed to be zero. Additional computations show that corrections to include a transverse velocity distribution can be incorporated in accord with reference \cite{focus2}. That is, for a Gaussian transverse velocity distribution peaked about zero and of width $\sigma$, we find that the deposited peaks are broadened by $\approx \sqrt{2}t_{int}\sigma$ while the ratios of peak height to background level are decreased by $\approx \frac{1}{\sqrt{\pi}t_{int}\sigma}$, where $t_{int}$ is the interaction time between the molecule and the field. 

Classical trajectories are computed for the motion of the $N_2$ center of mass in the presence of $V(x)$, which is encountered for the time period $t=0$ to $t=t_{int}$\cite{footnote}. We adopt an aspect of the ballistic aggregation model\cite{ba} and assume that all molecules that strike the surface stick without diffusing. Note also that although trajectories are computed for $N_2$ as a point particle, the $V(x)$ encountered by $N_2$ depends on the molecule's $J,M$ through its effect on $\chi$\cite{footnote2}.

\begin{figure}[!h]
\epsfxsize=6.0in
\hspace*{0.0cm}\epsffile{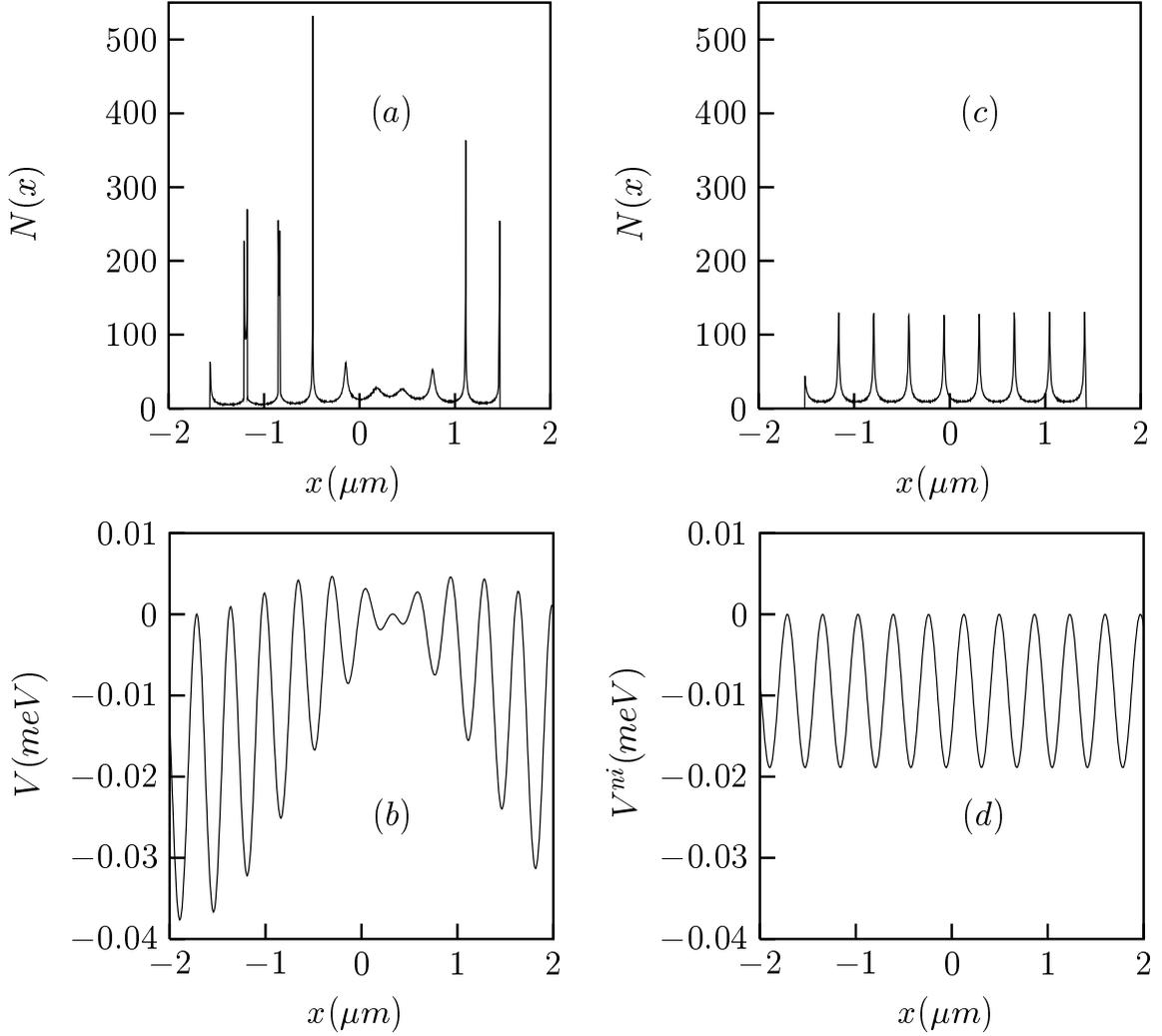}
\vspace*{-5.0cm}\caption{Molecular deposition and associated optical potential  for the initial superposition $\sqrt{0.2} \ket{0,0,0} +\sqrt{0.8} \ket{0,2,0}$ due to $V(x)=V^{in}(x) + V^{ni}(x)$ [panels (a) and (b)], and due to $V(x)^{in}$only. Here $\frac{E_2^{0}}{E_1^{0}}=1.0 \times 10^4$, $E_1^{0}=1.0\times10^2$
V/cm, $\lambda   _1=0.628 \mu m$,  $\lambda _2=0.736  \mu  m$, $\theta_F=-2.65$ radian, and $t_{int}=0.625~ \mu$ sec. Results are from a sample of 20,000 trajectories.}
\label{fig2}
\end{figure}

Consider first simple cases involving only a single superposition of states. Figure \ref{fig2}  shows the results in the presence and absence of interference contributions for a superposition comprised of $\ket{0,0,0}$ and $\ket{0,2,0}$. Specifically, panels (a) and (b) show the pattern of deposition, and the associated optical potential, for dynamics in the presence of $V(x)=V^{in}(x) + V^{ni}(x)$. For comparison we show, in panels (c) and (d), the corresponding results assuming that there is no coherence between $\ket{\phi_1}$ and $\ket{\phi_2}$, i.e. neglecting  $V^{in}(x)$. In absence of molecular coherence the optical potential is seen to be (panel d) periodic, resulting in a series of short periodic deposition peaks (panel c). By contrast, the inclusion of interference contributions (panels a and b) result in significant enhancement and narrowing of peaks [ five times narrower (FWHM of less than 4 nm) and four times more intense ] as well as the appearance of an aperiodic potential and associated aperiodic deposition pattern.

Quantitative consideration of the peaks shows that they are in general accord with the theory outlined in Ref. \cite{focus2}. That is, a sharp peak forms in the region of the potential minima  when $t_{int}\sim(2n+1)T/4$, where $T$  is the optical period for a particular potential  well. In the presence of $V^{in}(x)$ not all potential wells have the same period. Hence, deposition is not periodic and is dependent on the interrelationship between $t_{int}$ and the period $T$ of each different well.

The optical potential $V(x)$ and hence the nature of the deposition pattern, is seen to depend analytically [see Eqs. (\ref{pottot})-(\ref{potin})] on the contributing $\ket{\phi_i}$, the coefficients $c_i$, the phase $\theta_F$, the fields $E_i^{(0)}$, and the  time of interaction $t_{int}$ between the field and the molecule. Of these, numerical studies on the relative phase $\theta$ of the $c_i$, an important parameter in coherent control studies of photodissociation and bimolecular scattering\cite{cc}, show that it does not significantly affect the deposition pattern.

\begin{figure}[!h]
\epsfxsize=6.0in
\hspace*{-2.0cm}\epsffile{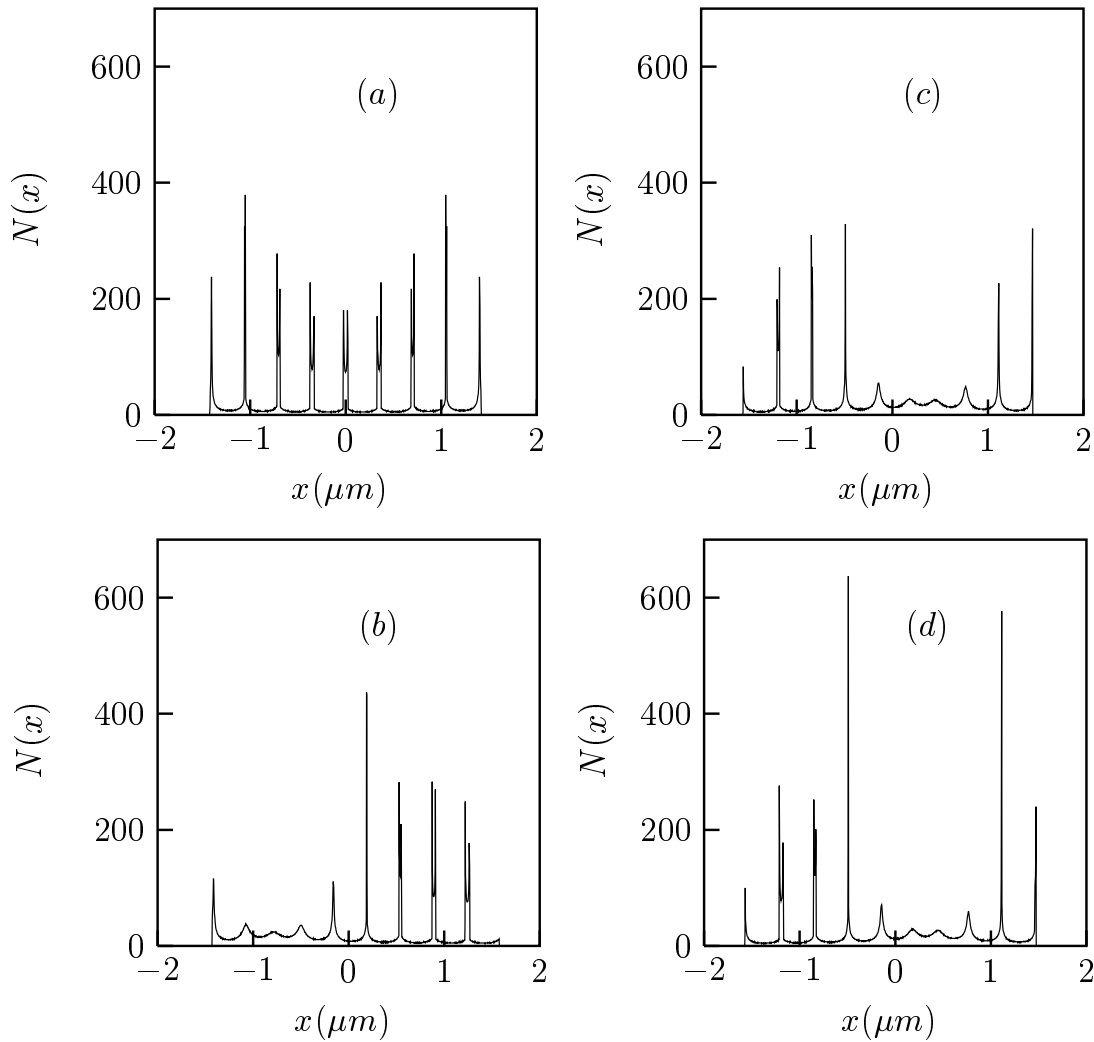}
\vspace*{-5.0cm}\caption{(a) and (b): Molecular deposition associated with $\sqrt(0.8) \ket{000} + \sqrt(0.2)\ket{1,2,0}$ for varying $\theta_F$; i.e. (a) $\theta_F = 0$, and (b) $\theta_F=2.0$. (c) and (d): Sample variation of deposition with changes in $|c_1|, |c_2|$: (c) $[1-\sqrt(0.01)] \ket{0,0,0} + \sqrt(0.01) \ket{0,2,0}$ and (d) $\sqrt(0.4) \ket{0,0,0} + [1-\sqrt(0.4)] \ket{0,2,0}$. Other parameters are as in Fig. \ref{fig2}.}
\label{fig3}
\end{figure}

Consideration of Eq. (\ref{pottot}) shows that this is because changes in $\theta$ do not affect the positions of the extrema of $V(x)$, and only result in small changes in the depth of the minima. By contrast, changes in the other parameters can strongly affect the structure of the deposited pattern. For example, panels (a) and (b) of Fig. \ref{fig3} shows significant differences in both the position and intensity of the peaks as a function of $\theta_F$. By contrast, consideration of the analogous plot  where only $V^{ni}$ is considered (not shown) shows no variation in peak intensity as a function of $\theta_F$. Similarly, panels (c) and (d) show the strong dependence of the deposition upon the magnitude of the coefficients of the created superposition. Clearly, varying these parameters affords a wide range of control over the deposited pattern. 

Finally, consider control over a beam of molecules with a thermal distribution of molecular level populations. That is, consider the case where the initial collimated molecular beam is in a mixture of states $\sum_{i,j} w_{i,j} \ket{0,J_i,M_j}\bra{0,J_i,M_j}$, with the weights $w_{ij}$ given by a Boltzmann distribution at $T=298^o$ K. In this instance 20 $J$ states are populated. By passing this mixture through a square pulse of field strength 3.25 $\times 10^9$ V/m and frequency width 75.4 cm$^{-1}$ we excite all nineteen states to pairwise superpositions of  $J$ states. That is, we produce the mixture $\sum_{i,j} w_{i,j} [~~c_{i,j}\ket{0,J_i,M_j}\bra{0,J_i,M_j} + d_{i,j}\ket{0,J_i+2,M_j}\bra{0,J_i+2,M_j}~~]$ where $|d_{i,j}|^2 =1 - |c_{i,j}|^2$.

\begin{figure}[!h]
\epsfxsize=6.0in
\hspace*{-2.0cm}\epsffile{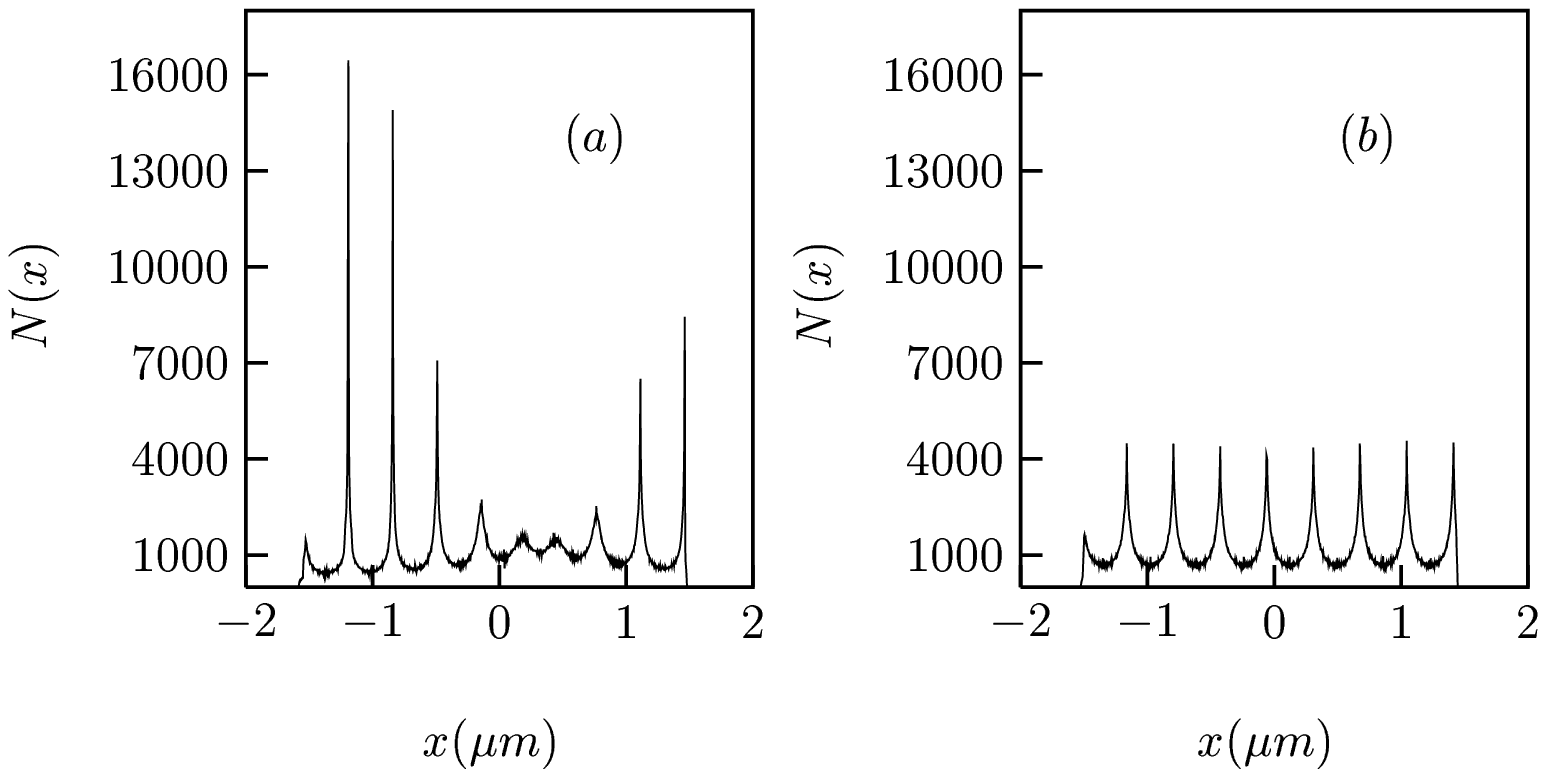}
\vspace*{-11.0cm}\caption{Molecular deposition  with (right)  and without (left)    molecular  coherence     for      the   mixed state at temperature $T=298^o$ K, as described in text. Remaining parameters are as in Figure \ref{fig2} except that results are obtained from 1.23 $\times 10^6$ trajectories and $t_{int} = 0.467 \mu m$.}
\label{fig4}
\end{figure}

At the chosen field strength, $d_{ij}$ can be computed in perturbation theory\cite{mccullough}, the final result being that the mixture of superpositions has the  coefficient $c_{0,0}$ associated with the state $\ket{0,0,0}$ on the order of $\sqrt{0.8}$. This mixed state is then passed through the two stationary fields and the deposition pattern computed. Results for one such case are shown in Fig. \ref{fig4} where results including the coherence contributions $V^{in}$ are shown in panel (a) and contrasted with the results where only the non-interference terms are included [ panel (b) ]. The results are quite similar to those of the single superposition shown in Figures \ref{fig1}-\ref{fig3} above. That is, including the interference, in addition to eliminating the periodicity, results in more  intense,  sharper  lines.  Examination  of $V(x)$ as a function of $J,M$ shows that the lack of broadening of the peaks with mixing of $J,M$ levels is a result of the fact that changing $J,M$ alters only the depth of the $V(x)$ minima, and not their location. 

In summary, we have shown that introducing a coherence between molecular energy levels, in conjunction with two frequency related electromagnetic fields, introduces a set of parameters that allow for control over the nanoscale molecular deposition pattern. Further work is needed to consider the possibility of depositing any arbitrary pattern, to examine the focusing of larger  molecules (which have inherently larger polarizabilities and should be more easily focused) and to consider the effects of more intense CW fields. Work to this effect is in progress.

{\bf Acknowledgement :} Support from the  U.S. of Naval Research is gratefully acknowledged.

\begin{references}
\bibitem{footnote} For simplicity in the computation, we do not include a free flight time from the nozzle to the onset of the field in the computation. Including this time would increase both the number of molecules lost from the target area due to the transverse velocity spread as well as the width of the individual peaks.
\bibitem{footnote2} Thus, for example, molecules may well arrive at the surface with some degree of alignment, since molecules with different $M$ are subject to different forces.
\end{references}
\end{document}